A Comparative Case Study on the Impact of Test-Driven Development on Program Design and Test Coverage
By Siniaalto, M., & Abrahamsson P

This is the author's version of the work.

# A Comparative Case Study on the Impact of Test-Driven Development on Program Design and Test Coverage

Maria Siniaalto and Pekka Abrahamsson
*VTT Technical Research Centre of Finland*
*P.O.Box 1100, FIN-90571 Oulu, Finland*
*{firstname.lastname}@vtt.fi*

**Abstract**

*Test-driven development (TDD) is a programming technique in which the tests are written prior to the source code. It is proposed that TDD is one of the most fundamental practices enabling the development of software in an agile and iterative manner. Both the literature and practice suggest that TDD practice yields several benefits. Essentially, it is claimed that TDD leads to an improved software design, which has a dramatic impact on the maintainability and further development of the system. The impact of TDD on program design has seldom come under the researchers' focus. This paper reports the results from a comparative case study of three software development projects where the effect of TDD on program design was measured using object-oriented metrics. The results show that the effect of TDD on program design was not as evident as expected, but the test coverage was significantly superior to iterative test-last development.*

## 1. Introduction

Test-driven development (TDD) is not, despite its name, a testing technique but rather a development technique in which the tests are written prior to the source code [1]. The tests are added gradually during the implementation process and when the tests are passed, the code is refactored to improve its internal structure. This incremental cycle is repeated until all the functionality is implemented. [2]

The idea of TDD was popularized by Beck [3] in the Extreme Programming (XP) method. Therefore, although TDD seems to have just recently emerged, it has existed for decades; an early reference to the use of TDD features in the NASA Project Mercury in the 1960s [4].

Both the literature and practice indicate that the use of TDD yields several benefits. Among others, TDD leads to improved test coverage [2] and simplifies the design by producing loosely coupled and highly cohesive systems [5]. TDD also enables the implementation scope to be more explicit [5]. As a positive side effect, TDD may lead to enhanced job satisfaction and confidence [5]. Also, larger teams of developers can work on the same code base because of the more frequent integration [1]. On the other hand, TDD has received criticism for not compensating the lack of up-front design and not being very suitable for systems such as security software or multithreaded applications, since it cannot mechanically demonstrate that their goals have been met [6]. It is also claimed that rapid changes may cause expensive breakage in tests and that the lack of application or testing skills may produce inadequate test coverage [7]. However, the scientific empirical evidence behind these claims is currently scattered and disorganized, and thus it is difficult to draw meaningful conclusions. Moreover, studies addressing the impact of TDD on program design are currently very scarce. To our knowledge, there are a few studies in which the quality of the code design has been studied, but only one has really addressed the impact of TDD on program design [8].

The purpose of this paper is twofold. First, it reviews the empirical body of evidence and organizes it systematically. Second, the paper presents the results from a comparative case study on the impact of TDD on program design in semi-industry settings by comparing a TDD project with two iterative test-last projects using a suite of agreed object-oriented metrics.

The explicit focus of this empirical study is to investigate whether TDD improves design quality and test-coverage as claimed by its proponents. The object-oriented (OO) metrics used for inspecting the design quality are as follows: weighted methods per class (WMC), depth of inheritance tree (DIT), number of children (NOC), coupling between objects (CBO), response for a class (RFC) [25] and lack of cohesion in methods (LCOM*)[30].

The results partially contradict the current literature as the empirical evidence shows that TDD does not improve the program design as expected. In particular, TDD does not necessarily produce highly cohesive systems when employed by non-professional developers. However, TDD appears to result in significantly better test coverage than the iterative test-last development. The findings are based on three software development projects, two of which used iterative test-last development and one utilized TDD.

## 2. Review of the Empirical Body of Evidence

In this section, the existing empirical evidence on TDD is reviewed and summarized. A total of sixteen studies were included in this review. The main results of each study are collected in a table in Appendix 1.

### 2.1. Classification of TDD studies

The studies were categorized by their context: four of the studies [9-12] were classified as "performed in an industrial context" since they were conducted primarily with professional developers in real industrial settings developing a concrete software product to be delivered. Five studies [8, 13-16], which were either conducted with professional developers implementing exercises designed for a particular experiment or with students working in close-to-industry settings aiming to deliver an industrial software product , were classified as "performed in a semi-industrial context". The remaining studies [17-23] were categorized as "performed in an academic context" because their test subjects were mostly undergraduates employing TDD in a set of experimental tasks. Table 1 presents the TDD study classification schema used to organize the reviewed studies.

**Table 1. TDD study classification schema**

| Type | Subjects | Context | # studies |
|---|---|---|---|
| Industry | Industrial developers | Real project | 4 |
| Semi-industry | Industrial developers | Experimental task | 4 |
| | Student developers | Real project | 1 |
| Academic | Student developers | Experimental task | 7 |
| | | Total | 16 |

### 2.2. TDD in Industrial Settings

Bhat and Nagappan [9] present the results of two industrial case studies conducted at Microsoft. They compared the impacts of TDD and non-TDD development processes on quality and overall development time. Their results show that the quality of the code developed using TDD increased 2.6–4.2 times when compared to non-TDD developed code. Alternatively, the project managers estimated that TDD increased the overall development time by 15–35 %. The block coverage was 79–88 % at unit test level in projects employing TDD. They also noticed that the tests serve as auto documentation when the code was maintained or used.

Lui and Chan [10] present the results of TDD and software process improvement in China. The authors find that TDD improves task estimation as well as process tracking. TDD also enhances the following consistent practices and guidelines. Their results show that the defect rate was significantly decreased in teams employing TDD. TDD teams were also able to fix their defects faster.

Williams et al [24] and Maximillien and Williams [11] present the findings of a case study of industrial programmers at IBM. They notice a significant defect rate reduction in the new project using TDD; approximately 50 % when compared to the third release [11] and 40 % when compared to the seventh release [24] of an old project. They also observe that the developers felt that the daily integration of TDD saved them from late integration problems as these results were gained with minimal impact on developer productivity.

Damm at al. [12] present the experiences of industrial developers using a customized test automation tool with a component level approach to TDD. Their initial project evaluations indicated significantly decreased fault rates and project lead-time.

As the most important finding, the defect density was reduced in all the studies conducted in industrial settings. There were also indications that TDD may improve test coverage [9], accelerate defect fixing and enhance estimation and process tracking [10]. However, the observed productivity effects were contradictory, as in one study TDD shortened the overall project lead time [12], but in two studies the productivity results were reduced [9] or slightly reduced [11, 24].

### 2.3. TDD in Semi-Industrial Settings

Canfora et al. [13] arranged a controlled experiment with professional developers to compare TDD and iterative test-last development in terms of productivity and the quality of the unit tests. The experiment consisted of two 5 hour runs. The test subjects implemented every assignment using both development approaches, but in different runs. The results suggest that TDD supports better performance predictability and may improve unit testing, but it also requires more time.

Müller [8] studied the effect of test-driven development on program code. He included five TDD software systems of which three were student projects and compared them with three open source based conventional software systems. He developed a new metric called assignment controllability (AC) to investigate the effect of test-driven development on the resulting code. Müller used Chidamber and Kemerer's [25] object-oriented metric suite to measure the program code, which is also used in this study. He observed that OO metrics did not show any impact on the use of TDD but that the new metric, AC, was able to show a difference. More specifically, he identified that the number of methods where all assignments are completely controllable is higher for systems developed with TDD. Müller also found some evidence of testability increase in TDD projects.

George and Williams [14] performed controlled structural experimental trials with professional programmers to compare TDD with the conventional test-last development. Their experimental findings indicate that the TDD practice yields code with improved external code quality: the TDD pairs passed 18 % more test cases. The productivity was found slightly reduced as it took 16 % more time to complete the assignments. However, the comparison of the productivity is somewhat uneven, because the authors noticed that the control group tended to leave the test cases unwritten. They also report that TDD test cases achieved a mean of 98 % method, 92 % statement and 97 % branch coverage.

Geras et al. [15] conducted an experiment to measure the differences of TDD and traditional test-last development. The experiment involved experienced programmers as research subjects. They noticed no significant differences in developer productivity, although TDD seemed to produce more tests and the TDD practitioners ran the tests more frequently. The incidence of failures at the acceptance test level appeared to be greater when using TDD, but at the unit test level that was not an issue.

Abrahamsson et al. [16] report the results of a controlled case study on TDD approach in a mobile Java environment. The developer team consisted of three students and a professional developer. They report that the developers did not see the benefit of writing tests and indicated a strong reluctance to adopt TDD. The authors suggest that this was due to the application domain not being very suitable for TDD and the developers would have needed constant mentoring during the project. The amount of test code was 7.8 % of the total lines of code, and the total percentage of time used in test code development was 5.6 %.

The quality effects of TDD are not as obvious in this context as in the studies performed in industrial settings. However, there are indications that TDD may help to produce more tests [15] and to achieve better test coverage [14]. On the other hand, TDD may result in a false sense of security and therefore cause more failures at the acceptance test level. [15] The adoption of TDD may also be difficult and it is not necessarily suitable for all application domains [16].

### 2.4. TDD in Academic Settings

Janzen and Saiedian [17] arranged a formal experiment to study the effects of TDD on programmer productivity, test coverage and internal quality. They used upper-level undergraduate students as research subjects. The subjects were divided into three teams: the first team used TDD while the second one used iterative test-last development and the third team only performed manual testing after the implementation phase. The results indicate that TDD may have a positive impact on developer productivity. The test coverage was relatively low regardless of the development approach. They calculated several structural and OO metrics to evaluate any possible difference in internal quality. Although most of these results were within acceptable limits, there were some concerns regarding complexity and coupling in TDD code and manually tested code. The study also pointed out that the developers perceive TDD more positively after they have tried it.

Kaufmann and Janzen [18] present the results of a controlled experiment conducted with undergraduate Computer Science students comparing TDD with test-last approach. Their results showed that the TDD group was significantly more productive producing 50 % more code than the control group. They found no differences in the code complexities between the groups. However, there were indications that the design quality of the TDD group was better and their programmer confidence was noticeable higher. They also report that these findings may be the result of greater programming experience among the subjects practicing TDD.

Müller and Hagner [19] analyzed a controlled experiment to compare TDD with traditional test-last development with graduate Computer Science students as test subjects. Their experiment did not show any significant differences between the approaches in terms of the development time. They notice that TDD provides programmers with an improved program understanding and therefore they are able to use existing methods faster, correctly. The reliability of the final programs was measured using random and acceptance tests. The random tests did not reveal any notable differences, but the acceptance tests showed that the final reliability of the TDD program was significantly lower.

Panþur et al. [20] ran a controlled experiment studying TDD and iterative test-last development. The experiment was conducted with 38 senior undergraduate Computer Science students. Their findings did not show any significant differences between the examined approaches, although TDD seemed to slightly reduce the external code quality and average mean of code coverage. The majority of their test subjects also felt that adopting TDD was difficult and that TDD was not very effective.

Erdogmus et al. [21] compared TDD with iterative test-last development. Their experiment involved undergraduates as test subjects. They observe that TDD appears to improve productivity. It also seems to improve the understanding of the requirements and to encourage better decomposition. The small scope of the tests and more frequent feedback reduced debugging and rework effort. They found no differences in quality, but TDD seemed to produce more consistent quality results. They also observed that the minimum quality increased linearly with the number of tests, independent of the development technique used.

Steinberg [22] presents the findings of the use of unit testing in an XP study group consisting of faculty members and some upper level undergraduates. Although, his work concentrates on discussing on the benefits of this phenomenon from the educational point of view, it also provides information about the effects of TDD and is thereby included in this study. His results indicate that TDD facilitates correcting the faults. The students also tended to write more cohesive code when using TDD and the coupling was found to be looser, since the objects had more clearly defined responsibilities.

Edwards [23] presents the experiences of the use of TDD with an automated grading system in a classroom. A total of 59 Computer Science students participated in a course that utilized TDD. The course assignments were the same as previously when also 59 students completed the course without using TDD. Both assignments were analyzed using the same automated grading tool. According to the analyses, the students using TDD and the automated grading tool produced code with 45 % less defects. Their developer confidence was also improved.

The findings in academic settings are partly contradictory: there are indications that TDD may improve software quality significantly [23]. However, in [20] the external code quality slightly decreased. In two studies [18, 22] program design improved with TDD, whereas there were some concerns regarding the design's complexity in [17]. The results also indicate that TDD may facilitate implementation work [21, 22] and improve developer confidence [18, 23], although in [20] the developers considered TDD ineffective. All studies addressing TDD's productivity issues observed positive effects.

### 2.5. Summary

While many of the findings point in a positive direction, their generalizability and significance remain questionable. One reason for this is the fact that often the metrics used for describing the findings have not been either defined in detail or lack the quality attribute they should be presenting. Another element limiting the industrial applicability of these study results is the scope of the tasks carried out within the experiments: real software products often consist of several thousands of lines of code and require several developers' work contribution. However, quite many of the experiments consisted of several short tasks and were as small as only a few hundred lines of code.

Half of the studies included in this review were performed in the academic context and therefore the external validity could be limited. However, this is arguable because the studies comparing students and professionals have concluded that similar improvement trends can be identified among both groups [26] and that the students may provide an adequate model for the professionals [27].

Based on the findings of the existing studies, it can be concluded that TDD may improve software quality significantly and enhance and facilitate the work of the developers. The productivity effects of TDD are not that obvious, and the results vary regardless of the context of the study. However, there are indications that TDD does not necessarily decrease developer productivity or extend project lead-times.

## 3. Empirical Results from a Comparative Case Study

TDD directs the construction of code's structure and affects the way of implementing the code as the tests are written prior to the source code. Hence, its impacts should be visible from the very start. The main goal of this comparative empirical evaluation of TDD is to explore the impact of TDD on program design and test coverage. This section begins with the layout of the research design for the study. Then, the empirical results are presented and discussed and study limitations are identified.

### 3.1. Research Design

The research method in the three case projects is the controlled case study approach [28]. It combines aspects of experiments, case studies and action research and is especially suitable for studying agile methodologies. The approach involves conducting a project which has the business priority of delivering a specific software product to a customer, in close-to-industry settings, where measurement data is constantly collected.

All three projects aim to deliver a concrete software product to a real customer. The projects were not simultaneous. The implementations were realized with Java programming language. The project teams worked in a shared co-located development environment during the projects. All projects continued for nine weeks, comprised of six iterations and followed an agile software development method, Mobile-D™ [29], which is an empirically developed method for the purposes of the research context. It embodies a defined daily development rhythm and improvement mechanisms for systematic software process improvement. For the purposes of this study, the method provides a coherent framework for studying iterative test-last and TDD.

Table 2 provides a summary of the parts of the projects which are not convergent to each other. The development teams in the projects consisted of undergraduates with 5-6 years of completed studies. All the team members in projects 1 and 2, which used iterative test-last development, had some industrial coding experience whereas only one developer in project 3, which used TDD, had previously worked in industrial settings. However, all the subjects in project 3 were Software Production majors and had a personal interest in programming. The developers were told and encouraged to write tests in all projects regardless of the development technique used. In addition, in project 3, the use of TDD was stated mandatory. Intranet applications were implemented in projects 1 and 3: in project 1 for managing research data and in project 3 for project management purposes. Both systems consisted of a server side and graphical user interface. A stock market browsing system to be used via a mobile device was implemented in project 2. The biggest part was handled by the server side and the mobile part mainly focused on connecting the server and retrieving data. However, to make the comparison of TDD and iterative test-last even, graphical user interfaces and the mobile client application part in project 2 were excluded from the evaluation. The difficulty of implementation was at the same level in all projects.

**Table 2. Summary of the case projects**

|  | Project 1 | Project 2 | Project 3 |
|---|---|---|---|
| # of developers | 4 | 5 | 4 |
| Development technique | Iterative test-last | Iterative test-last | TDD |
| Product type | Intranet app. | Mobile app. | Intranet app. |
| Product concept | Research data management | Stock market browser | Project management tool |
| Product size (LOC) | 7700 | 7000 | 5800 |

### 3.2. Results

The object-oriented design metrics for measuring the design quality effects of TDD within this study were selected from the ones proposed by Chidamber and Kemerer [25]: weighted methods per class (WMC), depth of inheritance tree (DIT), number of children (NOC), coupling between objects (CBO) and response for a class (RFC). The lack of cohesion in methods (LCOM) was, however, replaced with the LCOM* proposed by Henderson-Sellers [30] because the original LCOM does not always provide valid results [31]. Table 3 presents the metrics that were used to investigate the impact of TDD on design quality.

**Table 3. Used metrics**

| Metric | Object Oriented Construct | Source |
|---|---|---|
| WMC | Complexity | [25] |
| DIT | Inheritance | |
| NOC | Inheritance | |
| CBO | Coupling | |
| RFC | Collaboration / Complexity | |
| LCOM* | Cohesion | [30] |

The test coverage was measured at method, statement and branch levels. The method coverage measures the ratio of executed methods and all possible executable methods during the testing. Statement coverage measures whether each possible statement is executed and branch coverage measures which possible branches in the flow control structures are executed. The subsequent sections present the empirical results of each metric.

**3.2.1. Coupling between Object Classes.** CBO [25] presents the number of classes to which the class is coupled. CBO can be used as an indicator of whether the class hierarchy is losing its integrity, since high coupling between object classes means that modules depend on each other too much and will be hard to reuse. Independent classes are easier to reuse in other applications. In order to improve modularity and promote reuse, inter-object class couples should be kept to a minimum. High coupling also makes the maintenance more difficult, because the sensitivity to changes is higher as well. Figure 1 presents the coupling results of the case projects. As there are indications that the TDD code may be less coupled, it should also be noticed that the TDD results are the most dispersed and the coupling is quite low in all three cases. Whether these small differences are explained by the use of a particular development technique cannot be concluded.

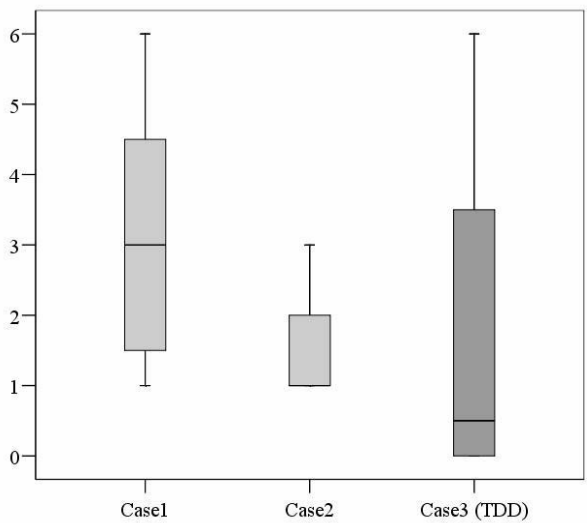

**Figure 1. Coupling between object classes.**

**3.2.2. Lack of Cohesion in Methods.** LCOM* [30] measures the correlation between the methods and the local instance variables of a class. High cohesion indicates good class subdivision while lack of cohesion increases complexity. This algorithm produces answers in the range 0 to 1, with the value zero representing perfect cohesion and with value one presenting extreme lack of cohesion.

Figure 2 presents the LCOM* results of the case projects. Project 3, in which TDD was used, has clearly the worst cohesion. It is also noticeable that the results are the most consistent in project 3, which probably indicates that the poor cohesion is not likely to be a coincidence. Whether it is due to TDD or the developers' lack of experience remains to be analyzed. However, this result already indicates that TDD does not automatically produce highly cohesive code.

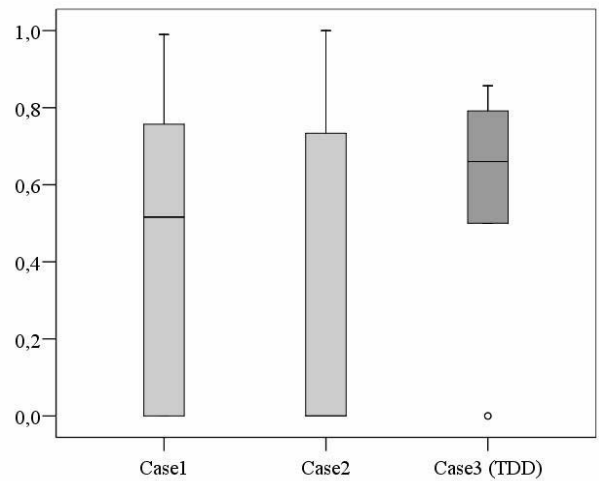

**Figure 2. Lack of cohesion in methods.**

**3.2.3 Other Object-Oriented Metrics.** Other OO-metrics were analyzed as well. However, as they did not show any noticeable differences between the used development techniques, they are not included here.

**3.2.4. Test Coverage.** Figure 3 presents the method, statement and branch coverage of the projects. All the results were significantly better in project 3 which employed TDD. This is notable, because all the developers, regardless of the development technique used, were encouraged to write tests. In projects 1 and 2, the test classes were implemented, they were not as thorough as the ones implemented in project 3. It should be pointed out that the use of TDD and achieved better test coverage did not affect the developer productivity as the aimed product was delivered on time.

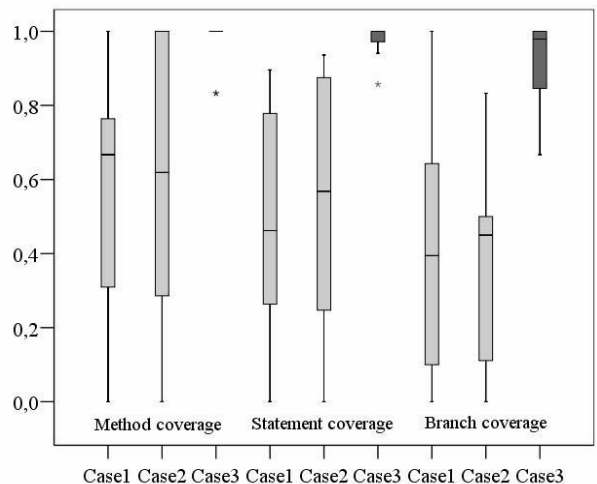

**Figure3. Test coverage.**

**3.2.5. Summary of Empirical Findings**. In the empirical evaluation, the design quality effects of TDD were explored through six OO-metrics: WMC, DIT, NOC, CBO, RFC and LCOM*. The evaluation was performed by calculating the values of the metrics using the data obtained from the three case projects introduced in section 3.1. The results of WMC, DIT, NOC and RFC did not reveal any significant differences between TDD and iterative test-last development. Based on the results of these four metrics, all projects had an acceptable design.

The value of CBO indicates that TDD produces slightly less coupled code. Nonetheless, the differences between the projects are small and the coupling is low in all three projects regardless of the development technique used. Therefore final conclusions cannot be drawn in any direction. The value of LCOM* provided contrasting results: the code produced in the TDD project (project 3) was significantly less cohesive than code produced in projects 1 or 2, in which iterative test-last development was employed. Especially significant was the fact that the LCOM* results were clearly most consistent in the TDD project. This finding is also contradictory with the claims presented in the literature and it indicates that TDD may require a certain level of professional expertise that was not evidently present in the TDD project. Therefore, it can be concluded that the impact of TDD on program design is not necessarily a positive one in the hands of less experienced developers. To be precise, TDD does not automatically result in highly cohesive program code. These initial findings support the findings of Müller [8] who confirmed that the OO-metrics did not cause a difference in his study either.

The test coverage was assessed at method, statement and branch coverage levels. The empirical data produced by these three projects clearly indicate that the test coverage of the TDD project can be very high without sacrificing the productivity, as the aimed end product was delivered to the customer as planned. These results are in line with the existing literature.

**3.3. Threats to Validity**

The variety of the programming experiences in the case projects pose a threat to the internal validity of this study. In industry, teams usually have a mixed set of experience and skills, as many students work in industrial companies before their graduation. When doing industrial case studies, all these developers are usually regarded as professionals. On the other hand, it can be the case that a student with a strong personal interest in programming is more skilled, even though he does not have any industrial experience. It is possible that the developers in the TDD project were not as skilled as the developers in the iterative test-last projects. However, as all of the developers were undergraduates and none of them was employed by any industrial company, we regard them all as non-professionals.

The product types and concepts were different in all the projects. To make the comparison fair, we excluded graphical user interfaces and the mobile client application part in case 2. We also believe that all product implementations presented a similar level of difficulty.

The subjects' conformance to correct implementation of the tests and use of TDD was controlled by assigning a person responsible for testing in all projects. The amount of test code was monitored in all projects. The significant differences in test coverage between the development techniques were not due to a slip in the given instructions. The test classes were implemented in all projects, they were more thorough in the TDD project. Possible observation effects were limited, as the developers were not aware that the program design or the test coverage was to be compared. The measurements were accomplished after the project endings.

The use of students as study subjects places a question about the generalizability of the results. However, the development environment was designed to relate closely to industrial development setting with strict deadlines, regular working hours and most importantly, the developers were developing a real product for real customers. In this respect, it has been observed that similar improvement trends can be identified among students and professionals [26], and that students may provide an adequate model for professionals [27].

Another threat to the external validity arises from the size of the software product as well as the distribution of the project work. All the studied projects had less than 10 000 lines of code, their development took around 1000 person hours and the products were developed by a single team in one location. The impact of TDD on program design, however, should be visible from the very start and thereby present in the studied projects.

# 4. Conclusions

Test-driven development is one of the most fundamental practices of agile development. It is said to yield several benefits, yet some of these claims have not been empirically studied or supported. Especially, the claim that TDD leads to improved software design has not been widely addressed.

This research has two main contributions: first, it reviews and organizes the existing empirical body of evidence regarding the impact of test-driven development on software development. Secondly, it presents the results of the impact of TDD on program design as well as on test coverage.

The design metrics listed in this paper provide valuable data for other researchers investigating TDD either in academic or industrial settings. Practitioners will find it useful to correlate the case metrics to their systems in particular in web-based and mobile application areas.

Interestingly, some of the results obtained in this study contradict some claims in the literature. Here, the empirical data indicates that TDD does not always produce highly cohesive code as suggested in the literature. This is the case, at least, when the TDD users are inexperienced developers. As a positive contribution, the test coverage is, however, significantly improved. This is an important finding since it strengthens the empirical understanding of a promising programming technique. Furthermore, the results presented in this paper are important as they contribute to the gradual build up of empirical evidence on software engineering innovations.

**Appendix 1. Empirical body of evidence on TDD**

| [1] | # of subjects | Study focus/ Comparison [2] | Empirical findings (+ positive, - negative, * neutral) | ref |
|---|---|---|---|---|
| Industrial | 2 projects, 11-14 dev. | Non-TDD | + Code quality was 2.6–4.2 times better while the block coverage was 79-88 %. The tests also served as documentation.<br>- The project managers estimated that the overall development time increased by 15-35 %. | [9] |
| Industrial | 2 teams | Traditional test-last | + Task estimation and process tracking improved, along with the following of practices and rules. The defect density reduced significantly and the developers were able to fix defects faster. | [10] |
| Industrial | 9 dev. | Ad-hoc unit testing | + Defect density was reduced by 40–50 %. The developers felt that the daily integration diminishes integration problems.<br>- Developer productivity was minimally impacted. | [11, 24] |
| Industrial | 2 projects | Specialized testing tool | + Defect density and project lead-time reduced significantly. | [12] |
| Semi-industrial | 28 dev. | Traditional test-last | + TDD supports better performance predictability and may improve unit testing.<br>- TDD extends lead time. | [13] |
| Semi-industrial | 5 TDD and 3 convent. projects | Traditional projects | + TDD may increase testability.<br>* Muller developed a new metric called assignment controllability, which is higher with TDD code. Other OO metrics did not show any difference. | [8] |
| Semi-industrial | 24 dev. | Traditional test-last | + TDD pairs passed 18 % more test cases. The test cases achieved 98% method, 92% statement and 97% branch coverage.<br>- The productivity was reduced as it took 16 % more time to complete the assignments. | [14] |
| Semi-industrial | 14 dev. | Traditional test-last | + TDD seemed to produce more test cases and the tests were run more frequently.<br>- The incidence of failures at the acceptance test level was higher with TDD. | [15] |
| Semi-industrial | 4 dev. | Exploratory data | - Developers indicated a strong reluctance to adopt TDD and would have needed constant mentoring during the project.<br>* TDD may not be suitable for all application domains | [16] |
| Academic | 3 teams | Iterative test-last, No tests | + TDD may have a positive impact on developer productivity.<br>- There were some concerns regarding complexity and coupling in TDD code.<br>* The developers perceived TDD more positively after they had tried it. | [17] |
| Academic | 8 dev. | Test-last | + TDD group produced 50% more code and their programmer confidence was noticeably higher. There were also indications that the design quality of the TDD group was better.<br>* They found no differences in the code complexity. | [18] |
| Academic | 19 dev. | Traditional test-last | + TDD promotes better program understanding and the developers were able to use existing methods faster correctly.<br>- The final reliability of the TDD code was significantly lower at the acceptance test level. | [19] |
| Academic | 38 dev. | Iterative test-last | -TDD slightly decreased external code quality and the average mean of code coverage. The majority of the test subjects felt that adopting TDD was difficult and that TDD was not very effective. | [20] |
| Academic | 24 dev. | Iterative test-last | + TDD appeared to improve productivity and understanding of the requirements. It also encouraged better decomposition and reduced debugging and rework effort.<br>* TDD produced more consistent quality results. The minimum quality increased linearly with the number of the tests, independent of the development technique used. | [21] |
| Academic | Not known | Exploratory data | + TDD facilitated the correcting of the faults. The students wrote more cohesive code with looser coupling with TDD. | [22] |

---

[1] Study classification as presented in section 2.1.
[2] In some studies, the compared technique was not specified in detail.